\documentclass[endfloats*,showkeys,preprintnumbers,preprint,amsmath,amssymb,epsf,jcp]{revtex4}
\usepackage{dcolumn}
\usepackage{epsfig}

\usepackage{graphicx}
\usepackage{longtable}
\usepackage{dcolumn}
\usepackage{bm}

\def\Xa{X$\alpha$}
\def\af{$\alpha$}
\def\afHF{$\alpha_{HF}$}
\def\afEA{$\alpha_{EA}$}
\def\trip_b{$^3$B$_1$ }
\def\sing_a{$^1$A$_1$ }
\def\af{$\alpha$ }
\def\afHF{$\alpha_{HF}$ }
\def\afOPT{$\alpha_{OPT}$ }
\def\Xa{X$\alpha$ }
\def\rhup{\rho_{\uparrow}}
\def\rhdn{\rho_{\downarrow}}
\def\vrp{\vec{r'}}
\def\vr{\vec{r}}
\def\bo{\overline{\rho}^{\frac{1}{3}}}
\def\ro{\overline{\rho}}
\def\bt{\overline{\rho}^{\frac{2}{3}}}
\def\ss{\sigma}
\def\bo{\overline{\rho}^{\frac{1}{3}}}
\def\ro{\overline{\rho}}
\def\bt{\overline{\rho}^{\frac{2}{3}}}
\def\gbo{\overline{g}^{\frac{1}{3}}}

\def\gbt{\overline{g}^{\frac{2}{3}}}
\def\ss{\sigma}

\begin{document}
\preprint{NRL-GWU/008}

\title{The limitations of Slater's element-dependent exchange functional from analytic density 
functional theory}


\author{Rajendra R. Zope}
\email{rzope@alchemy.nrl.navy.mil}
\affiliation{Department of Chemistry, George Washington University, Washington DC, 20052, USA}
\altaffiliation{Mailing address: Theoretical Chemistry Section,   Naval Research
Laboratory, Washington DC 20375-5345, USA}

\author{Brett I. Dunlap}
\email{dunlap@nrl.navy.mil}
\affiliation{Code 6189, Theoretical Chemistry Section, US Naval Research Laboratory
Washington, DC 20375}

\date{\today}

\date{\today}

\begin{abstract}
     Our recent formulation of the analytic and variational Slater-Roothaan
(SR) method, which
uses Gaussian basis sets to variationally 
express the molecular orbitals, electron density and the one body effective
potential of density functional theory, is reviewed. 
Variational fitting can be extended to the resolution of identity method,
where variationality then refers to the error in each two electron 
integral and not to the total energy. However, a Taylor series analysis 
shows that all analytic {\it ab initio} energies calculated with variational
fits to two-electron integrals are stationary.
It is proposed that the appropriate fitting functions be charge neutral 
and that all {\it ab initio} energies be evaluated using two-center 
fits of the two-electron integrals.
The SR method has its root 
in the Slater's  \Xa method and permits an arbitrary scaling of the 
Slater-G\`asp\`ar-Kohn-Sham exchange-correlation potential around each 
atom in the system. The scaling factors are the Slater's exchange 
parameters $\alpha$. Of several ways of choosing these parameters, two 
most obvious are the Hartree-Fock (HF) \afHF values and the exact atomic \afEA values. 
The former are obtained by equating the self-consistent \Xa energy and 
the HF energies, 
while the latter set reproduce {\em exact} atomic energies. In this work, 
we examine the performance of the SR  method  for predicting
atomization energies, bond distances, and ionization potentials using the
two sets of \af parameters.  The atomization energies are calculated for 
the extended G2 set of 148 molecules for different basis set combinations.  
The mean error (ME) and mean absolute error (MAE) in atomization energies 
are about 25 and 33 kcal/mol, respectively for the exact atomic,  \afEA, \,
values.  The HF values of exchange parameters, \afHF, give 
somewhat better performance for the atomization energies with ME 
and MAE being about 15 and 26 kcal/mol, respectively.
While both sets give performance better 
than the local density approximation or the HF theory, the errors
in atomization energy are larger than the target chemical accuracy. 
To further improve the performance of the SR method for atomization 
energies, a new set of \af values is determined by minimizing the 
MAE in atomization energies of 148 molecules. This new set gives
atomization energies half as large (MAE $\sim$ 14.5 kcal/mol)
and  that are slightly better
than those obtained by one of the most widely used generalized 
gradient approximations. Further improvements in atomization energies require
going beyond Slater's element-dependent functional form for exchange employed 
in this work to allow exchange-correlation interactions between electrons 
of different spin.
The MAE in ionization potentials
of 49 atoms and molecules is about 0.5 eV and that in bond distances of 27 molecules 
is about 0.02 \AA.   
The overall good performance of the computationally efficient SR method 
using any reasonable set of $\alpha$ values makes it a promising method 
for study of large systems.

\end{abstract}

\pacs{ }


\maketitle
\newpage

   Theoretical methods for electronic structure calculations in practice 
today can be broadly classified into two categories. 
From the perspective of density functional theory (DFT), these two classes of models 
differ from each other only in the way
the unknown exact exchange-correlation (XC) energy functional is approximated.
The first class of methods are the 
traditional quantum chemical methods such as Hartree-Fock (HF) theory and beyond
\cite{Pople99}.  These methods are generally implemented using Gaussian 
basis sets. The molecular orbitals are analytically expressed as linear combinations of 
atomic orbitals (LCAO). The atomic orbitals are contracted sets of primitive
Gaussian basis functions.  One advantage of the Gaussian basis set is that it permits
computation of matrix elements of the energy and a number of other properties 
completely analytically, which allow them to be computed to whatever precision is desired.
Another advantage of Gaussians is that they are very localized in real and momentum space
resulting in sparse matrices.  The primary disadvantage of Gaussian-based
HF is that it formally scales as $N^4$, where $N$ is the number of atomic orbital basis functions.
Despite that single disadvantage HF is often the choice of users for geometry-optimization
of molecules because usually some combination of basis-set choice, cutoff and other
approximations define
HF calculations that are the fastest all-electron calculations using current, commercial, quantum
chemistry computer packages.  HF-based methods can alway be systematically improved towards the
exact results.

Density functional (DF) based models \cite{HK,KS65} form the second class
of methods for studying the ground-state properties of materials. 
The most important aspect of density functional theory is relative computational simplicity
which leads to better scaling and the ability to optimize the functional form of the Kohn-Sham (KS)
multiplicative potential.  Today, KS density functional models are the predominant choice for
pure electronic structure calculations.  DFT formally scales as $N^3$.
Various implementations of the DF models that use a variety of different 
type basis sets as well as those that are fully numerical (use no basis sets) 
have been reported. Most of these implementations including those that use 
analytic Gaussian basis sets require use of numerical mesh to compute the 
contribution from the XC term. The complexity of the functional forms, particularly
for all-electron calculations using the gradient-corrected functionals can
require a very sophisticated numerical grid of points to integrate the XC energy density
to get the total energy accurately.
The use of grids makes calculations of matrix elements that are accurate 
to arbitrary precision (as in the Hartree-Fock method) practically 
impossible and leads to number of problems such as steps or kinks 
in the potential energy surface, spurious negative frequencies, etc. 
Also, a numerical total energy incorrectly depends on
how the molecule is oriented with respect to the orientation of the numerical grid.
Consequently, the total energy is not exactly rotationally invariant
\cite{BJohnson,Dunlap03}.  Some of these 
problems can be eliminated by using very refined numerical grids
which, needless to say, leads to reduction in computational speed
that compromises the inherent efficiency of the DF models.  DF 
implementations that are fully analytic are obviously desirable. 

While translating the Slater-type-orbital DF method of Baerends, Ellis and
Ros \cite{Baerends}
for Gaussian-type-orbitals, Sambe and Felton \cite{Sambe} proposed treating the
KS equation completely analytically, which was a significant first
step towards analytic DFT, however a numerical grid was needed to
fit the XC potential.  This works quite well for the subset of DF theory
 in which the same fits
can be used in the KS equations and the energy thereby almost preserving
the variational principle; i.e. preserving the variation
principle when a complete basis set is used to fit the XC potential
\cite{Dunlap79}.
These efforts greatly improved upon the muffin-tin (MT) approach to DFT, which spherically
averaged the KS potential introducing a discontinuity in the potential
at the surface of the spheres.  Direct integration of the potential in 
the MT approximation leads
to an undefinable energy.  Instead, as is common
even today with diverse formulations of DFT, one simply makes the same
approximations in evaluating the energy that one makes in computing the
potential \cite{MT1,MT2,MT3,MT4}.  In such cases today one 
argues, where possible,
that the energy and the KS potential are computed exactly, therefore the
calculation is variational.  In this work the calculation of the energy
and potential is inexact due to incomplete basis sets, but the
calculations are to machine-precision explicitly variational.

Our variational solution \cite{Dunlap79} to the problem of fitting
any charge distribution has become quite popular, particularly for simplifying MP2.
While adding second derivatives \cite{Kormornicki} to a descendant of our code, DGauss
\cite{AW91}, a different fit was used to simplify the MP2 energy expression.
That new fit was called Resolution of the Identity (RI)\cite{RI}. 
 In a later application of their RI method, one  of the definers of RI
showed empirically that our fitting method was better than the original
approach\cite{Vahtras}.
Many have attributed this improvement to some unknown \cite{Vahtras,Eichkorn}
or not clearly relevant \cite{Ahlrichs} properties of the Coulomb norm and have told us
that others \cite{Bloor,Whitten} used the Coulomb norm before our work.  
Our work concerns variational fitting, whereas RI, which if it is not 
precisely our method, requires  a complete basis set.

    The fits used in the later RI work are precisely the variational
solutions that maximize the unique robust (no first-order error due to fitting)
 Coulomb self-energy of the fitted charge 
distributions. 
 Furthermore,
with these fits, the RI energy calculated using three-center approximations
to the two-electron integrals is identical to the same expression using the
two-center approximations to the two-electron integrals that are obtained using 
fits to both charge distributions\cite{Dunlap00a}.  In the following, we examine
 a Taylor-series expansion
in the error made in fitting two-electron integrals in all 
{\it ab initio} energies.
We define good (but not perfect) fits to have negligible quadratic and higher errors.
All XC-related, robust and variational fits used in this work do not involve 
the Coulomb norm.

Chemists are beginning to say that our method
does not work in RI applications to large systems \cite{Varga,Jung_HG,Gill}.
The physics problem is that an unbalanced charge, no matter how small or 
how far separated, when
arranged on a infinite, periodic lattice has an infinite Coulomb self-energy.
A way to handle this fact is to constrain the fits to have the
right charge \cite{Baerends,Mintmire82,Mintmire83,Trickey}.  If this is done, then
by Gauss' law the electrostatic potentials of the fitted and exact charge
distributions become identical as soon as they possibly can,  outside the two
distributions.  Our code uses both constrained and unconstrained fits.  The fits
are robust, and if they are good (but not perfect) then 
the constrained and unconstrained energy are the same within typical quantum 
chemical tolerances.

This work also uses robust and variational fits of the XC potential.  
A set of DFT models that have the same functional form in the energy density 
and the KS potentials,
$$ E_{xc} [\rho] =   \frac{3}{4}  \int \rho(\vec{r}) v_x[\rho(\vec{r}),\vec{r}]d^3r,$$
with
\begin{equation}
 v_x [\rho] =  - \alpha \frac{3}{2}
 \Bigl( \frac{3}{\pi} \Bigr )^{1/3} \rho^{1/3}(\vec{r}) ,
\end{equation}
where $\alpha$ is the Slater exchange parameter (which was found 
to be $1$ by Slater\cite{Slater51} and $2/3$ by G\`asp\`ar\cite{Gaspar}  and Kohn
and Sham \cite{KS65} (GKS)),  is easiest to treat analytically.
The difference in the two values of $\alpha$ 
has roots in the averaging process employed in the simplification of the HF  
exchange potential. Slater obtained $\alpha=1$ while simplifying the nonlocal
exchange potential of the Hartree-Fock approximation 
by averaging exchange potential over 
the entire Fermi sphere of radius $k_f = (3\pi^2\rho(\vec{r}))$. 
 The GKS value of $2/3$ is obtained by applying the variational 
method to the statistical total energy expression, and only uses an average 
over surface of the Fermi sphere $k = k_f(\vec{r})^{1/3} $.
Later, Kohn and Sham also set  $\alpha$ to be $2/3$ for general 
energy functionals through the construct of the non-interacting electron gas. 
Most DF models that are in practice 
today use the GKS value of $\alpha$. Recently, overall improvement in 
performance of some of these DF models has been noted if $\alpha$ is allowed 
to vary\cite{ZD_CPL04,OLYP,O3LYP}.

   In the \Xa method, the form of the exchange potential is given by Eq. [1] and 
the \af are the scaling parameters.  
 The \Xa method is the outcome of  simplifying the Hartree-Fock 
method.  It was suggested that the scaling parameters \af could be 
obtained by ensuring that the self-consistent \Xa energy for atoms matches 
the Hartree-Fock energy. Such set of \af parameters was determined by 
Schwarz \cite{Schwarz72}. Subsequently, several other ways to determine the \af 
parameters  were 
proposed \cite{Smith70,Smith78,a1,a2,a3,a4,a5,OOS2000,ZD_JCTC,Fliszar1,Fliszar2,Fliszar3,Fliszar4}.
The muffin-tin implementation of the \Xa method allowed
the use of atom-dependent $\alpha$ values.  The unique advantage of this
quantum-chemical method was that the model smoothly dissociates into atoms as it was
pulled apart, independent of the basis-set (partial waves in the muffin-tin
method).  With the ability to choose the atomic energies however one wanted,
it is a way to extrapolate atomic properties to all homogeneous and
heterogeneous materials, or alternatively the properties of all elemental
molecules or crystals to all heterogeneous materials.

A fully analytic method that allows arbitrary 
scaling of exchange potential around each type of atom in the heteroatomic 
system was recently formulated  and implemented\cite{Dunlap03}. 
It is called Slater-Roothaan (SR) method.  It is free from the problematic
energy of the multiple 
scattering \Xa model but retains all its advantages.
It is computational very efficient and has been successfully applied to study boron 
and aluminum nitride nanotubes contains about two hundred atoms\cite{BN,AlN}. 

The accuracy of the SR method ultimately depends on the choice of scaling parameters, $\alpha$'s,
used in calculations. 
Two obvious choices for choosing $\alpha$ are: (1) the \af$=$\afHF values that give the HF
energy for atoms, and  (2) the \af$=$\afEA values that give the {\em exact} atomic energies.
The use of later set is appealing because when used in molecular calculations, as the 
molecule dissociates  the corresponding sum of the exact atomic energies can be obtained.  
Our early calculations on the total energies of molecules shows that remarkably 
accurate total molecular energies can be obtained using the second set\cite{ZD_PRB05}.
These molecular energies of the G2 set of 56 molecules are better than 
or comparable to most pure or hybrid density functional models (such as 
PBE\cite{PBE}, B3LYP\cite{B3LYP} etc,
see  Ref. \onlinecite{ZD_PRB05} for details). 
In the present work, we examine the overall performance of the Slater-Roothaan method 
for the atomization energies, bond distances, and ionization potentials when the 
above sets of \af values are used. We use the extended G2 set of 148 molecules 
for the benchmark. This set is routinely used to examine the performance of 
various density functional and related models.  As we shall see, the performance
for atomization energies with these two combinations of Gaussian basis sets, 
though better than the local 
density approximation or the Hartree-Fock method, is inferior to that 
of the DFT models with generalized-gradient approximations (GGA).  Many 
density functional models are parametrized to give better atomization 
energy. Following this practice, we explore the possibility of improving the
performance of Slater-Roothaan method for atomization energies. 
This is accomplished by 
determining a new set of \af values that minimize the mean absolute error (MAE) in the atomization
energies of 148 molecules.  This parametrization gives accuracy comparable to some of 
widely used gradient-corrected density functional models.

\section{Theoretical Method}
\subsection{Variation and the Eigenvalue Problem}

    Our interest is in the development of analytic DFT through robust and variational fitting.  The
total electronic energy in Hohenberg-Kohn-Sham \cite{HK,KS65} DFT for an $N$-electron
system is a functional of electron density $\rho$. The electron density is given by
\begin{equation}
      \rho_{\sigma}(\vr) = \sum^N_{i} n_{i,\sigma} \phi_{i,\sigma}^*(\vr) \phi_{i,\sigma}(\vr),
\end{equation}
where, ${\phi_{i,\sigma}(\vr)}$ are single particle orbitals, and 
 $n_{i,\sigma}$, the occupation numbers 
for both spins. The total electron density $\rho$ is the sum of spin 
densities $\rhup$ and $\rhdn$. In chemistry, the orbitals are still usually expressed in LCAO form,
\begin{equation}
      \phi_{i,\sigma}(\vr) = \sum_{j} C_{ij} u_{j}(\vr) 
\end{equation}
The energy is determined by constrained variation of the energy with respect to orthonormal orbitals, 
\begin{equation}
      {dE\over dC_{ij}} = \epsilon_{ik} \langle \phi_{k} u_j \rangle ,
\end{equation}
where the angular braces indicate the overlap integral and the $\epsilon$ matrix is all the Lagrange
multipliers needed to allow the orbitals to be made orthonormal while minimizing the energy.
In DFT a unitary transformation can
always be found to diagonalize this matrix to generate a conventional molecular-orbital eigenvalue problem.
This eigenvalue problem is called the Kohn-Sham (KS) equations.  We approximate these equations
by fitting in uniques ways to make them less computationally challenging while
preserving the  full variational principle.

The first robust and variational fit was to the self-Coulomb energy of a charge distribution,
\begin{equation}
  E_{ee} = \langle  \rho\vert\vert\rho \rangle  =
    \frac{1}{2} 
 \int  \int \frac{\rho(\vr)\rho(\vrp)}{\vert\vr-\vrp\vert} d^3r \, d^3r'.
\end{equation}
This energy is  approximated  by expressing the  charge density as a fit to a set of Gaussian
functions,
\begin{equation}
\rho(\vr) \approx \overline{\rho}(\vr) = \sum_i d_i G_i(\vr),
               \label{Eq:D}
\end{equation}
where, $\ro (\vr)$ is the fitted density, d$_i$ is the expansion coefficient of the charge 
density Gaussian basis-function G$_i$. The elimination of the first order error in the total energy due to the fit
leads to the unique robust expression for the self-Coulomb energy\cite{Dunlap79}
\begin{equation}
  E_{ee} \simeq \overline{ E_{ee}} \equiv 2 \langle \ro\vert\vert\rho \rangle  -  
            \langle \ro\vert\vert{\ro} \rangle = \langle \ro\vert\vert\rho \rangle + \
	    \langle \Delta\rho\vert\vert\rho \rangle,
               \label{Eq:E_c}
\end{equation}
where, $\Delta \rho = \rho-\ro$. Unconstrained variation of this energy gives 
\begin{equation}
   \vec{d} = \langle \vec{G} \vert\vert \vec{G}\rangle^{-1} {\bf \cdot}
             \langle \vec{G} \vert\vert \rho\rangle .
                 \label{Eq:Dv}
\end{equation}
In this work, an overline represents any approximation, good, bad or exact.  
Thus the identity is not necessarily resolved, and we do not use the RI method \cite{RI,Vahtras}.
Hopefully we and others can generate basis sets that deliver good fits.  

  In any event we single out fitted
energies that contain no first-order error and call them robust, because 
mathematically speaking, it makes no sense
to include non-robust, approximate  energies in any eigenvalue 
problem.  The fit,
if it is good, will change if the original charge distribution does; the derivatives of 
good fitting coefficients with respect to LCAO orbital coefficients are not zero. 
In the derivative of the total energy, however, this derivative is multiplied by zero 
if the fit is obtained by variation of a robust energy.
It is easy to obtain atomic density-fitting bases for
which the first-order error is small and approximately equal to the square root of the total (quadratic)
error made in this approximation \cite{MD}. (Of course, variational fits only have 
a first-order error if they are constrained.)

With the advent of fitting basis sets that are individually optimized for corresponding orbital basis sets
\cite{GSAW92,EWTR97}, it is appropriate to define a stronger definition of good fit.
%
%
If we can use basis sets that are accurate to one part in 10$^{-5}$ then 
a robust fitted energy is accurate to 
 is about one
part in 10$^{-10}$,
which is a appropriate number to be used for molecular integral cutoffs, defining convergence, etc.,
in {\it ab initio} quantum chemistry codes
and can probably be safely neglected in most of quantum chemistry.
We call a fit this good or better a good fit.  
Such a fit is only  a good fit, we do not claim it to be exact.
This is our target, which at this point we cannot guarantee.

One could, of course, use a non-robust expression for the fitted energy and then correct the effects of that
error on the orbital variational problem by taking the total derivative of the energy, as written in Eq. 3.
In that case the corresponding Fock matrix or KS equations differ from what would be obtained by inserting
the non-robust fit where appropriate in the fit-free eigenvalue or KS equations.  The potential of the
eigenvalue problem cannot be made independent of the energy.  If the
dependence of the energy on the orbitals is not treated variationally, then the forces are not accurate
\cite{Andzelm}.  This fundamental problem probably means that the proposal of Sambe and Felton \cite{Sambe} 
to fit the exchange-correlation (XC) potential of DFT to LCAO form by numerical sampling (or direct numerical
integration) must be abandoned.  Now almost all DFT codes treat the XC potential by direct numerical
integration, despite the fact that exact forces arising from using an auxiliary basis to numerically
fit the XC potential are known \cite{Andzelm2} and have been implemented precisely for gradient corrected
functionals \cite{Dunlap97}.  The exact partitioning scheme of Becke \cite{Becke87} together with
adaptive grids that vary with the orbital basis set \cite{Pederson} are now overwhelmingly popular with LCAO
approaches to DFT \cite{Koster04}.  Intermediate in the trade-off between accuracy and efficiency between
the Sambe-Felton and Becke approaches is one in which
the variationally-fitted density (Eq. (\ref{Eq:D}) ) replaces the two-center exact density in the numerical construction of
the XC matrix elements \cite{Koster04a}.

\subsection{Robust Fitting and the Resolution of the Identity}

The post-HF {\it ab initio} energies are functions of the two-electron integrals
\begin{equation}
  \langle \phi_i\phi_j\vert\vert\phi_k\phi_l \rangle \equiv
  \langle ij\vert\vert kl \rangle .
       \label{Eq:IJKL}
\end{equation}
RI energies are defined to be functions of $\langle ij\vert\vert \overline{kl} \rangle$,
where $\overline{kl}$ is the fit to the $kl$ orbital pair.
The post-HF {\it ab initio} energies can be expanded in a Taylor series in the
difference between between an exact two-electron energy and any approximate two-electron energy,
\begin{equation}
  E(\langle ij\vert\vert kl \rangle)
  -E(\overline{\langle ij\vert\vert kl \rangle})
  \, = \sum_{n=1}^\infty\,  {1 \over n!}\, 
     {\partial^n E \over \partial \langle ij\vert\vert kl \rangle^n}\,  \Delta^n_{ijkl},
\end{equation}
where
\begin{equation}
  \Delta_{ijkl}= \langle ij\vert\vert kl \rangle-\overline{\langle ij\vert\vert kl \rangle}.
\end{equation}

The unique robust two-electron energy 
\begin{equation}
  \langle ij \vert \vert kl \rangle \approx \overline{\langle ij \vert \vert kl \rangle}_{rob}
  = \langle ij \vert \vert \overline{kl} \rangle  + \langle \overline{ij} \vert \vert kl \rangle 
 - \langle \overline{ij} \vert \vert \overline{kl} \rangle, 
   \label{Eq:robust}
\end{equation}
where both fits are separately determined by Eq. (\ref{Eq:Dv}),
is beginning to be used directly in {\it ab initio} quantum chemistry \cite{Manby}.  
No matter how both fits are obtained there is no first-order error,
\begin{equation}
  \langle ij \vert \vert kl \rangle - \overline{\langle ij \vert \vert kl \rangle}_{rob}
  = \langle \Delta_{ij} \vert \vert \Delta_{kl}\rangle.
\end{equation}
If Eq. (\ref{Eq:Dv}) is used then all approximations are exactly equal \cite{Dunlap00ab},
\begin{equation}
  \langle ij\vert\vert \overline{kl} \rangle = \langle \overline{ij}\vert\vert kl \rangle =
  \langle \overline{ij}\vert\vert \overline{kl} \rangle,
\end{equation}
and the RI and robust (Eq. (\ref{Eq:robust})) approximations to the two electron 
integral are identical.

    The Coulomb norm is not
magical.  Coulomb potentials from  charge distributions can be approximated robustly by fitting the potential
due to a charge distribution rather than fitting the charge distribution itself \cite{MD}.
Obtaining good basis sets, however, is more of a problem than in the charge distribution fitting case
\cite{Dunlap83}, which is being overcome \cite{Manby-Knowles}.   
For {\it ab initio} energies,
\begin{equation}
 E(\langle ij \vert \vert kl \rangle) - E(\overline{\langle ij \vert \vert kl \rangle}_{rob})
 = \sum_{n=1}^\infty {1 \over n!} {\partial^n E \over \partial \langle ij\vert\vert kl \rangle^n}
 \langle \Delta_{ij} \vert\vert \Delta_{kl} \rangle^n_{ijkl}.
\end{equation}
If both fits are good, then this equation practically equals zero, i.e,
 $\langle \Delta_{ij} \vert\vert \Delta_{kl} \rangle,$ being
quadratic in errors, is expected to be  accurate to one part in $10^{10}$.
  Thus, for good RI fits, the
fitted energy {\em is} practically identical to the original energy
\begin{equation}
  E(\langle ij \vert \vert kl \rangle) 
   \simeq E(\overline{\langle ij \vert \vert kl \rangle}_{rob}) 
   = E(\langle \overline{ij}\vert \vert kl \rangle)
   = E(\langle ij\vert \vert \overline{kl} \rangle)
   = E(\langle \overline{ij}\vert \vert \overline{kl} \rangle),
\end{equation}
where the last three equalities hold only if the fits are obtained using Eq. (\ref{Eq:Dv}).

These simplest, global fits are problematical for large clusters \cite{Varga,Jung_HG}.  They could be
constrained, in
which the case the robust, rather than either of these two approximate, two-electron energies 
that involve a single fit must be used
to obtain robust energies, independent of whether or not the Coulomb metric is used.
It might be better to consider changing the fitting basis.  The amount of charge
in the product of two molecular orbitals is either zero or one depending on whether or not the orbitals are
the same.  If the orbitals are the same, then our methods are sufficient \cite{Dunlap79,MD}. 
If the orbitals
differ, as is the case in MP2, then all fitting basis functions should 
contain no charge, and a similar $p$ RI basis for atoms corresponding to
our $s$-type density-fitting basis, would include all the atomic $p$ exponents in the orbital
basis, but rather than doubled simply incremented by the smallest $s$ exponent.
This picks up the
smallest $p$-type contribution possible from the product of primitive $p$ and $s$ orbital functions
and this basis has a flexibility equal to that of the $p$ orbital basis.  For most basis
sets this is smaller than the correspond $s$-type charge-density fitting basis.  
Since the basis is neutral and the product of different orbitals pairs to be fit is neutral,
then all interactions die off as fast as possible
by Gauss' law.  All non-$s$ fitting functions have no charge, therefore the $L > 1$ charge-density fitting
functions are probably a good basis sets for fitting both the diagonal and non diagonal products of two molecular
orbitals.  This is obviously a better basis than any currently used for studies of large systems.
 If the RI fitting basis behaves well asymptotically, then the Coulomb norm is likely to
be proved best again for standard RI approximations to the MP2 energy of infinite systems \cite{Raynor}.
A zero-charge basis will likely to be effective in density functional perturbation theory
as well, if the resolution of the identity\cite{casida,tddft2} or ideally robust and variational
fitting is used. With such a basis we envision keeping constrained fits as a simple 
option to test the stability of the calculation even for large systems.

\subsection{Variational Fitting}
We take the full variation of the energy with respect to the orbitals before solving the eigenvalue problem
in order to obtain precise forces.  If we modify the energy by adding  any number of robust fits,
\begin{eqnarray}
     f[a_i(\vr)] & = & f[\overline{a}_i(\vr)] + Order\big [(a_i-\overline{a}_i)^2\big], \\
     \overline{a}_i(\vr) & = &  \sum_{ij} a_{ij} A_{i}(\vr),
\end{eqnarray}
where $A_{i}(\vr)$ is an appropriate LCAO basis for the the i$^{th}$ fit, to approximate troublesome terms
in the energy,
then the Fock matrix for the corresponding eigenvalue equation is obtained by the chain rule of differentiation,
\begin{equation}
      {dE\over dC_{ij}} = \sum_{kl} {dE\over da_{kl}} {da_{kl}\over dC_{ij}}.
\end{equation}

If robust fits are available, then we can improve upon them by
obtaining the fits through a variation of our robust energy, exactly as orbitals are obtained.
If and only if the fits are variational can the same fits be present in both the energy and the
corresponding eigenvalue problem for the orbitals.

\subsection{Analytic formulation of the G\`asp\`ar-Kohn-Sham-Slater density functional model}
The KS energy is given by
\begin{equation}
E^{KS} [\rho]  = \sum_i^N <\phi_i| f_1| \phi_i> + E_{ee} + 
            E_{xc}\big [ \rhup, \rhdn \big] \label{eq:I}
\end{equation}
 where,  the first term contains the kinetic energy operator and the nuclear attractive potential due to
the $M$ nuclei,
\begin{equation}
f_1 = -\frac{\nabla^2}{2} - \sum_A^M \frac{Z_A}{\vert\vr - \vec{R_A}\vert}.
\end{equation}
The  second term in Eq. (\ref{eq:I}) represents the classical Coulomb interaction energy
of electrons discussed above.
It is approximated robustly through Eqs. (\ref{Eq:E_c}) and (\ref{Eq:Dv}).
 The last term E$_{xc}$ in Eq. (\ref{eq:I}) is the exchange energy, 
\begin{equation}
E_{xc} [\rhup, \rhdn]  =  - \frac{9}{8} \alpha \Big ( \frac{6}{\pi} \Big )^{1/3} \int d^3r 
            \Big [ \rhup^{\frac{4}{3}} (\vr) +  \rhdn^{\frac{4}{3}} (\vr) \Big]. \label{eq:Exc}
\end{equation}
  The form of above functional allows analytic calculations with the Gaussian basis to be performed.
For this purpose the one-third and two-third powers of the electron density are expanded in 
Gaussian basis sets: 
\begin{eqnarray}
            \rho^{\frac{1}{3}} (\vr) \approx \bo =  \sum_i e_i  E_i  \\
                                              \label{G2}
            \rho^{\frac{2}{3}} (\vr) \approx \bt =   \sum_i f_i  F_i .
                                              \label{G3}
\end{eqnarray}
 Here, $\{E_i\}$ and $\{F_i\}$ are independent Gaussian basis functions, while $e_i$ and $ f_i$ are expansion coefficients.
The exchange energy is then given by\cite{Dunlap86,Cook86,BID89,Cook95,Cook97}
\begin{equation}
E_{xc}   =  C_{\alpha} \Bigl [ \frac{4}{3} \langle \rho \,  \bo \rangle - \frac{
2}{3} \langle \bo  \,
     \bo  \, \bt \rangle
    + \frac{1}{3} \langle \bt \, \bt \rangle
 \Bigr ] ,
\end{equation}
 where $C_{\alpha} = - {9} \alpha \Big ( \frac{3}{\pi} \Big )^{1/3} .$
  Thus using the four LCGO basis sets (one for orbital expansion and three fitting basis sets) the 
total energy is calculated analytically.
The LCAO orbital coefficients and the vectors {\bf d}, {\bf e}, and {\bf f} are found by constrained
variation.

\subsection{Slater-Roothaan method}
   The expression for the total electronic energy in the Slater-Roothaan method has the 
following form:\cite{Dunlap03}
\begin{eqnarray}
  E^{SR} &  = & \sum_i <\phi_i| f_1| \phi_i> + 2 \langle \rho \vert\vert\ro\rangle
                - \langle \ro\vert\vert\ro\rangle    \nonumber \\
          & & \,\, - \sum_{\sigma = \uparrow, \downarrow}
             C_x \Bigl   [
            \frac{4}{3} \langle g_{\ss}  \, \gbo_{\!\ss} \rangle
           + \frac{2}{3} \langle {\gbo}_{\!\ss} \,  {\gbo}_{\!\ss}  \, {\gbt}_{\!\ss} \rangle
  \nonumber \\
         & & \,\, \, +  \,\frac{1}{3} \langle {\gbt}_{\!\ss} \,  \, {\gbt}_{\!\ss} \rangle 
    \Bigr ].
             \label{G4}
\end{eqnarray}
  Here, $C_x = C_{\alpha}/\alpha$; the partitioned $3/4$ power of the exchange 
energy density,
\begin{equation}
  g_{\ss} (\vr) = \sum_{ij} \alpha(i) \, \alpha(j) \, D_{ij}^{\ss} (\vr),
\end{equation}
where $D_{ij}^{\ss} (\vr)$ is the diagonal part of the spin density matrix, and the function,
\begin{equation}
   \alpha(i) =  \alpha_i^{3/8}
\end{equation}
contains the $\alpha_i$, the \af in the \Xa for the atom on which the atomic orbital $i$ is
the centered.  
The fits  to powers of $g_{\sigma}$ corresponding to Eq. 23-\ref{G3} are obtained 
variationally from Eq. (\ref{G4}).

\section{Computational details}

  The analytic SR method requires four Gaussian basis sets.  One for the orbital expansion 
and others to fit different powers of electron density, which we obtain from 
literature.
%
 We choose Pople's triple-$\zeta$ (TZ) 6-311G** basis\cite{O1,O2} and the DGauss\cite{AW91} valence
double-$\zeta$  (DZ) basis set\cite{GSAW92} called DZVP2 for orbitals basis sets.  
The {\sl s-}type fitting bases are obtained by scaling and uncontracting 
the {\sl s} part of 
the orbital basis.  The scaling factors are 2 for the density, 
$\frac{2}{3}$ for $\bo$ \,\, and $\frac{4}{3}$ for $\bt$.  These  scaled bases are used for 
all {\sl s-}type fitting bases.
Ahlrichs' group has generated a RI-J basis for fitting the charge density of a valence 
triple-$\zeta$ orbital basis set used in the {\sc Turbomole}
program \cite{EWTR97}.  The non-$s$ parts of Ahlrich's fitting bases are used in 
combination with 
the 6-311G**  orbital basis sets. Hereafter, we shall refer to this combination of 
basis sets as basis set {\bf I} or 6-311G**/RIJ . In combination with DZVP2 orbital 
basis, we use the {\sl pd} part of the  A2 charge density fitting basis. This 
will be referred to  as basis set {\bf II} or DZVP2/A2.  
The geometries of molecules were optimized using the Broyden-Fletcher-Goldfarb-Shanno (BFGS) 
algorithm\cite{BFGS1,BFGS2,BFGS3,BFGS4,BFGS5}.
The forces on atoms are rapidly computed non-recursively using the 4-j generalized Gaunt 
coefficients \cite{Dunlap02, Dunlap05}.   The atomic energies are obtained in the highest 
symmetry for which the self-consistent solutions have integral occupation numbers.
The atomization energy is computed from the total energy difference of optimized 
molecule and its constituent atoms.  The open shell atoms are treated 
in lower than spherical symmetry so that they could have integrals occupation
numbers. Thus, following point group symmetries were chosen: C$_{2v}$ for 
C, O, Si and S atoms, D$_{6h}$ for atoms F, Al, and Cl, D$_{3h}$ for B, and
$I_h$ for Li, N, Na  and P.
 
    A database of geometries of set of  148 molecules, known in literature as 
the extended G2 set, was built. The molecules were built with appropriate symmetries 
to expedite their structure optimization.  A new set of $\alpha$ values 
was  optimized by minimizing the mean absolute 
error in the atomization energies of the 148 molecules:
\begin{equation}
 \big \{\alpha_{OPT}\big\} = min \Bigg [ \sum_i^{148} \frac{\vert 
D_e^i(\{\alpha_j\})
 -D_e^i(exact)\vert}{148} \Bigg ],
\end{equation}
where, $D_e^i(\{\alpha_j\})$ is the atomization energy of the $i^{th}$ molecule with $j^{th}$ set of 
alpha values.
The $\{\alpha_j\}$ optimization was repeated by starting with two different 
sets of alpha values:
\afHF and \afEA. We used the simplex method\cite{BFGS5} and during each iteration 
of $\{\alpha_i\}$ optimization procedure, all 148 molecules were reoptimized.
The $\{\alpha_i\}$ optimization process all together involved about
10000 optimization of molecules, which was accomplished by the use of PERL scripts.
The final optimal set gives the best atomization energies for the simple 
element dependent Slater's exchange functional employed in the present model. It
is called \afOPT hereafter.

\begin{table}
\begin{ruledtabular}
\caption{The set of $\alpha$ values used in the present work.
The \afHF values are due to Schwarz \cite{Schwarz72} 
(spin-polarized by Connolly\cite{MT3}) and the 
\afEA values are from Ref. \cite{ZD_JCTC}. The \afOPT\, values in 
the last column are obtained by minimizing the mean absolute error 
in atomization energy of 148 molecules (see text for more details).}
\label{tab:alpha}
\begin{tabular}{lcccc}
 Molecule         &    \afHF & \afEA (6-311G**/RIJ) & \afEA (DZVP2/A2) & \afOPT \\
\hline 
 H   & 0.77627     &  0.777390   &  0.781240  & 0.753703  \\
 Li  & 0.77157	   &  0.791690   &  0.792110  & 0.839295  \\
 Be  & 0.76823	   &  0.795740   &  0.796140  & 0.557058  \\
 B   & 0.76206	   &  0.786750   &  0.786770  & 0.674430  \\
 C   & 0.75331	   &  0.776770   &  0.776650  & 0.675039  \\
 N   & 0.74522	   &  0.767470   &  0.767260  & 0.645482  \\
 O   & 0.74188	   &  0.765000   &  0.764480  & 0.657869  \\
 F   & 0.73587	   &  0.760660   &  0.760010  & 0.575118  \\
 Na  & 0.73115	   &  0.752040   &  0.752870  & 0.823779  \\
 Mg  & 0.72918	   &  0.749940   &  0.751200  & 0.768385  \\
 Al  & 0.72853	   &  0.748220   &  0.748690  & 0.743690  \\
 Si  & 0.72751	   &  0.745390   &  0.746020  & 0.767716  \\
 P   & 0.72620	   &  0.743240   &  0.743970  & 0.860982  \\
 S   & 0.72475	   &  0.742620   &  0.743500  & 0.743745  \\
 Cl  & 0.72325	   &  0.741970   &  0.742720  & 0.662447  \\
\end{tabular}
\end{ruledtabular}
\end{table}

\section{Results and Discussion}

 \subsection{Atomization energies}
   The two sets of \af values used calculations in these calculations are given in Table \ref{tab:alpha}. The last column in this table is  a new set of \af values, which as noted earlier, 
is parametrized to give the best atomization energies.
The computed atomization energies for the G2 set of 148 molecules are given in 
Table \ref{tab:AE}. The point group symmetries of molecules are also listed 
in the Table \ref{tab:AE}.
The last two rows in Table \ref{tab:AE} give the mean absolute error (MAE) and the 
mean error (ME) 
in the atomization energies of the 148 molecules. It is apparent from the MAE and ME that 
the basis sets effects are marginal. The errors with smaller basis DZVP2/A2 are of similar 
magnitude to those obtained by the larger 6-311G**/RIJ basis set.  For the exact atomic 
alpha, \afEA,
values the MAE is about 34 kcal/mol which is larger than that for the  atomic Hartree-Fock,
 \afHF,\, values ($\sim 
26 $ kcal.mol). In the former case, the atomic energies by construction have zero error. In the 
case of \afHF, the atomic energies are equal to the Hartree-Fock energies. So error in 
atomic energies is roughly the correlation energies of these atoms \cite{Lowdin}.
As the \afEA\, values give exact total energies of atoms, it was expected that their
use in molecular calculations will result in overall  improvement in molecular 
properties. This expectation is not unreasonable as atomic energies are exact by 
construction and the molecules in the dissociation limit would give sum of exact 
atomic energies. It is therefore surprising that the \afHF set performs better than the \afEA\, set.
The better performance for the former set is clearly 
a consequence of error cancellation of total energies of molecules and atoms. 
On the other hand, the MAE in atomization energies is the same as error in total energies 
when \afEA\, values of used. No error cancellation occurs in this case. Note that performance 
of many density functional models for atomization of energies is due to such cancellations
of errors in total atomic and molecular energies\cite{ZD_PRB05}. 
Many of these sophisticated functionals perform poorly for the total energies.
The ME reflect that use of \afEA\, and \afHF\, values lead to an overall overestimation of 
the atomization energies. The largest deviation of about 146 kcal/mol occurs 
for the C$_2$F$_4$ molecule. There are about 6 molecules with absolute errors 
larger than  100 kcal/mol. These are CF$_4$, CF$_3$CN, C$_6$H$_6$, C$_4$H$_4$O, C$_5$H$_5$N,
and C$_2$F$_4$.  
     
     Optimizing the SR method for atomization energies significantly improves its 
performance as can be seen from the Table \ref{tab:AE}. The MAE and ME in atomization 
energies are reduced to 14.5 kcal/mol and -4.5 kcal/mol.  The optimization process 
varies most the \af values of atoms on the left of periodic table
and those on the right side. The \af values of N, O, and  F are reduced with respect
to the \afHF and \afEA\, values. 
The maximum absolute error is also reduced 
significantly. 
Overall performance of \afOPT set for atomization energies is comparable
to that of the Perdew-Burke-Ernzerhof\cite{PBE} generalized gradient approximation
(PBE-GGA)\cite{AES00}.

\subsection{Bond distances}
    The bond distances of 27 selected diatomic molecules\cite{TVS98}
are compared in Table \ref{tab:BL}
with their exact counterparts. The MAE and ME in bond distances are in the last two 
rows of the same table. The maximum MAE is about 0.02 \AA.  Except for the larger 
basis with \afHF\, the ME are negative. 
Unlike LDA or GGA, there is no consistent trend of overestimation or underestimation. On the 
whole the bond distances are smaller than the experimental bond 
distances. This trend is different from that of the local density approximation (Slater exchange
+ Vosko-Wilk-Nusair correlation, called sometimes as SVWN) 
and GGA functionals\cite{BJohnson,Goddard_2004} 
which either overestimate or underestimate bond distances. 
The slight underestimation of bond distances, on average seems consistent 
with the on the whole overestimation of atomization energies. In particular,
the \afEA\, bond distances are shorter and the atomization energies higher.
  The present value of MAE (Table \ref{tab:BL} 
can be  reduced if molecules containing Li and Na are omitted. The largest deviation 
of 0.09 \AA\, occurs for the Na$_2$. Also, F$_2$ is another molecule which is difficult
for the present models. Indeed, our study on these diatomics show that even a very small 
value of \af leads to binding of the F$_2$ molecule\cite{ZD_JCTC}. Overall the bond distances
are fairly accurately predicted for  both the  \af values are used, with \afHF values 
giving somewhat better performance.  The bond distances with new \afOPT set are not given 
in Table \ref{tab:BL}. Using this set, the MAE in bond distances of 27 molecules increases 
to 0.04 \AA. 

\subsection{Ionization potentials}
   The ionization potentials of 49 atoms and  molecules\cite{Curtiss_IP}
 computed for the \afHF and \afEA\, values are 
compared with their experimental counterparts in Table \ref{tab:IP}. In computing the ionization 
potential (IP) the energy of the cation was calculated at the geometry of the neutral molecule. Also, 
the symmetry of the neutral system was assumed for  the cation. 
This choice of convenience  will tend to exaggerate the errors in IP for some systems.
For example, the IP of nitrogen atom and methane molecule, are overestimated. For N atom, 
broken-symmetry 
calculation will give lower energy for cation. Similarly, Jahn-Teller distortion will lower 
the energy of methane cation and will reduce the error in  the (adiabatic) IP. The MAE in 
IP for two sets 
of \af values are about 0.5 eV. Both, MAE and ME are of roughly similar magnitude for both 
basis sets combinations indicating that the basis set effects are negligible. The sign of deviations 
do not show any consistent trend of overestimation or underestimation.  
    The electron affinities could be similarly computed from the energy difference of 
the neutral molecule and its anion. However, we did not compute the electron 
affinities as the computation of electron affinities within the most density functional 
models (including the present one) is problematic as these quantities are sensitive
to the self-interaction of 
electrons. The presence of self-interaction error leads to exponential decay of the 
effective potential in the asymptotic region as opposed to the correct $-1/r$ 
asymptote.
The additional electron in the anion therefore experiences shallower potential than it
otherwise should. This sometimes leads to the positive eigenvalue for this electron 
although the total energy may be lowered. For this reason we did not compute electron 
affinities but we expect that accuracy of the present method for electron
affinity will be similar to that of the local density approximation.

\section{Summary}
   Analytic quantum mechanics using Gaussian basis sets has been extended.
 RI {\it ab-initio} energies evaluated using two electron integrals 
in which both charge distributions are variationally fitted are themselves 
variational with respect to the fitting parameters. The performance of 
fully analytic density functional model called the Slater-Roothaan method, 
 is examined for
various properties, using two different sets of \af values, for two different
basis sets. The comparison of two sets of \af values used in this work show
that \af values obtained by matching atomic Hartree-Fock energies (\afHF)
give better performance that those obtained by equating atomic energy to the
 exact atomic energies (\afHF). The basis set effects as judged from the
comparison of results obtained with two sets of bases show that smaller 
DZVP2/A2 basis provides accuracy comparable to 
the larger 6-311G**/RIJ basis.  The MAE in atomization energies 
of the SR method with \afHF and \afEA\, sets  are about 26 and 34 kcal/mol, 
respectively.  In comparison with the local density approximation 
or the Hartree-Fock method,
the SR  method with \afHF and \afEA\, sets performs quite well for 
atomization energies, bond distances and ionization potentials.  
Further improvement in performance of the Slater-Roothan method for 
atomization energies is obtained by parameterizing it for atomization 
energies.  
A new set of \af parameters (\afOPT),  is determined by minimizing the 
MAE in atomization energies of a set of 148 molecules.
With this set of \afOPT parameters, the mean and mean absolute
errors in atomization energies of 148 molecules are -4.5 kcal/mol and 14.5 
kcal/mol. This makes its performance for atomization energies comparable to 
that of the PBE-GGA\cite{AES00}.  
This is remarkable considering it is 
perhaps the simplest of the density functional models  
and it is certainly the oldest such model.
The hybrid and even more complex functionals like B3LYP\cite{B3LYP} 
or the PBE0\cite{PBE0} 
perform better than the PBE-GGA. To further improve the  performance of 
the SR method for the atomization energy 
it is necessary to go beyond the simple element dependent Slater's 
exchange functional form employed in this work.
Considering the overall performance of SR method,
it is a good starting point for optimization of large systems. Many calculations 
employ Hartree-Fock method to optimize geometries or to explore potential energy surfaces, 
and then compute properties of interest at a  more sophisticated level.  The analytic SR model 
scales better than the HF method and being analytic it is necessarily 
computationally efficient. 
One could use  the \af value used in this work to optimize structures and then perform 
single point calculations, if feasible, using more sophisticated models.  
The ability of density functional models to perform large scale 
calculations is well known. Of the plethora of density functional models of 
different accuracy, choosing the simpler ones like the local density approximation 
further enhances one's ability to study large systems and/or to perform 
longer molecular dynamics simulations. Indeed, several recent large scale simulations  are performed 
using local density approximation\cite{BBGZ01,PWGG04,ZKDZ04}. In a recent work, the 
local density approximation (SVWN) was shown to be about 55\% faster than the
B3LYP and 40\% faster than the BLYP for a calculation on 9-alanine 
system\cite{Merz05}. The analytic implementation like the present one, being
free from grids, will perform even better and is a promising alternative for large
scale calculations at modest accuracy.

     The Office of Naval Research, directly and through the Naval Research Laboratory,
and the
Department of Defense's  High Performance Computing Modernization Program, through the 
Common High Performance Computing Software Support Initiative Project MBD-5, 
supported this work.

\LTchunksize=170
\begingroup
\squeezetable
\begin{longtable}{lcdccccc}
\caption{The deviation in atomization energies for the set of 148 (extended G2 set) of molecules 
within different computational models- 
 M1:  SR-$\alpha_{EA}$/6311G**/RIJ. 
 M2:  SR-$\alpha_{EA}$/DGDZVP2/A2, 
 M3:  SR-$\alpha_{HF}$/6311G**/RIJ, 
 M4:  SR-$\alpha_{HF}$/DGDZVP2/A2, 
 M5:  SR-$\alpha_{Opt}$/DGDZVP2/A2, 
All energies are in kcal/mol and are calculated at the optimized geometries 
of molecules in the respective model. The exact atomization energies 
are from Ref. }
\label{tab:AE}
\\
\hline\hline
 Molecule         & Symmetry&   ~M1~ & ~M2~ & ~M3~ & ~M4~ & ~M5~ & {\em Exact} \\
\hline 
H$_2$&  D$_{6h}$ & -24.9& -22.7& -24.9& -22.8 & -23.4 & 110.0 \\ 
LiH&  C$_{6v}$ & -19.6& -23.9& -20.1& -24.4 & -19.6 &  57.7 \\ 
BeH&  C$_{6v}$ &   8.6& -16.9&   6.3& -18.5 &   1.7 &  49.6 \\ 
CH&  C$_{6v}$ & -16.7& -15.5& -17.1& -16.2 & -16.2 &  83.7 \\ 
CH$_2$($^3B_1$)&  C$_{2v}$ &   5.5&   7.2&   1.8&   2.6 &  -7.4 & 189.8 \\ 
CH$_2$($^1A_1$)&  C$_{2v}$ & -22.4& -19.1& -24.3& -21.7 & -25.6 & 180.5 \\ 
CH$_3$&  D$_{3h}$ &  -7.2&  -4.0& -11.5&  -9.4 & -19.5 & 306.4 \\ 
CH$_4$&  Td & -11.8&  -7.3& -16.9& -13.9 & -23.4 & 419.1 \\ 
NH&  C$_{6v}$ & -16.3& -14.6& -16.7& -15.4 & -15.1 &  83.4 \\ 
NH$_2$&  C$_{2v}$ & -24.4& -20.0& -25.5& -22.0 & -22.9 & 181.5 \\ 
NH$_3$&  C$_{3v}$ & -25.5& -18.9& -27.6& -22.6 & -25.8 & 297.3 \\ 
OH&  C$_{6v}$ &  -7.7&  -6.2&  -8.7&  -7.8 & -11.2 & 106.3 \\ 
H$_2$O&  C$_{2v}$ &  -4.1&   0.7&  -6.9&  -3.5 & -13.9 & 232.1 \\ 
HF&  C$_{6v}$ &   4.1&   7.7&   1.5&   4.3 & -10.7 & 140.7 \\ 
Li$_2$&  D$_{6h}$ & -17.8& -18.8& -17.8& -18.8 & -18.7 &  24.4 \\ 
LiF&  C$_{6v}$ &   8.9&  -2.4&   4.2&  -7.3 & -26.1 & 138.8 \\ 
C$_2$H$_2$&  D$_{6h}$ &  17.5&   8.9&  11.7&   1.1 & -13.1 & 405.3 \\ 
C$_2$H$_4$&  D$_{2h}$ &   9.2&  11.1&   0.6&   1.0 & -17.3 & 562.4 \\ 
C$_2$H$_6$&  D$_{3d}$ &   1.0&   7.4&  -9.3&  -4.6 & -19.8 & 710.7 \\ 
CN&  C$_{6v}$ &  11.5&   2.4&   8.3&  -0.3 &  -8.0 & 179.0 \\ 
HCN&  C$_{6v}$ &   1.4&  -8.5&  -3.3& -13.3 & -24.7 & 316.3 \\ 
CO&  C$_{6v}$ &  24.0&  11.3&  19.6&   7.4 &  -6.1 & 259.2 \\ 
HCO&  C$_{1h}$ &  29.1&  23.1&  23.4&  17.4 &   0.8 & 278.3 \\ 
H$_2$CO (formaldehyde)&  C$_{2v}$ &  21.0&  17.8&  14.5&  10.6 &  -8.3 & 373.4 \\ 
H$_3$CO$_{}$H &  C$_{1h}$ &  10.3&  15.4&   2.2&   5.7 & -13.5 & 511.6 \\ 
N$_2$&  D$_{6h}$ & -13.5& -24.9& -15.3& -26.2 & -32.6 & 228.5 \\ 
N$_2$H$_4$&  C$_{2}$ & -24.2& -12.9& -29.3& -18.8 & -15.2 & 437.8 \\ 
NO&  C$_{2v}$ &  10.3&   2.4&   8.2&   0.6 &  -6.3 & 152.9 \\ 
O$_2$&  D$_{6h}$ &  37.1&  33.6&  34.1&  30.7 &  20.1 & 120.4 \\ 
H$_2$O$_2$&  C$_{2h}$ &  14.7&  20.4&   9.8&  14.5 &  -1.0 & 268.6 \\ 
F$_2$&  D$_{6h}$ &  28.0&  29.8&  26.5&  27.8 &  13.7 &  38.5 \\ 
CO$_2$&  D$_{6h}$ &  67.6&  49.0&  58.1&  40.4 &  10.7 & 388.9 \\ 
SiH$_2$($^1A_1$)&  C$_{2v}$ & -26.0& -23.4& -25.1& -23.5 & -24.6 & 151.4 \\ 
SiH$_2$($^3B_1$)&  C$_{2v}$ &  -9.5&  -8.1&  -9.9&  -9.5 &  -6.9 & 130.7 \\ 
SiH$_3$&  D$_{3h}$ & -32.0& -30.4& -32.0& -31.9 & -29.0 & 226.7 \\ 
SiH$_4$&  Td & -39.7& -37.3& -39.0& -38.5 & -37.4 & 321.4 \\ 
PH$_2$&  C$_{2v}$ & -26.7& -22.4& -26.2& -22.3 & -10.2 & 152.8 \\ 
PH$_3$&  C$_{3v}$ & -38.6& -32.0& -38.1& -31.8 &   0.6 & 242.0 \\ 
H$_2$S&  C$_{2v}$ & -18.7& -13.0& -19.3& -14.0 & -14.4 & 182.3 \\ 
HCl&  C$_{6v}$ &  -4.8&  -3.7&  -5.7&  -4.9 &  -8.0 & 106.2 \\ 
Na$_2$&  D$_{6h}$ & -11.9& -11.6& -11.8& -11.5 & -11.9 &  16.8 \\ 
Si$_2$&  D$_{6h}$ &  -2.7&  -2.5&  -3.6&  -3.6 &  -1.2 &  74.7 \\ 
P$_2$&  D$_{6h}$ & -23.1& -22.7& -23.6& -23.2 & -18.7 & 117.2 \\ 
S$_2$&  D$_{6h}$ &   8.6&  10.1&   6.8&   8.1 &  10.2 & 101.6 \\ 
Cl$_2$&  D$_{6h}$ &   6.5&   8.0&   4.7&   6.0 &   0.3 &  57.9 \\ 
NaCl&  C$_{6v}$ &  -9.0&  -7.2& -11.0&  -9.7 &  -7.9 &  97.8 \\ 
SiO&  C$_{6v}$ &   7.8&   7.3&   4.0&   3.5 &  -2.5 & 191.2 \\ 
CS&  C$_{6v}$ &   8.0&   7.2&   4.8&   4.1 &  -0.9 & 171.2 \\ 
SO&  C$_{6v}$ &  17.5&  22.7&  14.1&  19.3 &   8.8 & 125.1 \\ 
ClO&  C$_{6v}$ &  15.2&  21.3&  12.8&  18.7 &   9.7 &  64.3 \\ 
ClF&  C$_{6v}$ &  19.1&  24.3&  16.2&  21.2 &   1.5 &  61.4 \\ 
Si$_2$H$_6$&  D$_{3d}$ & -54.1& -49.2& -54.4& -52.3 & -47.4 & 529.5 \\ 
CH$_3$Cl &  C$_{3v}$ &   6.9&  11.2&  -0.2&   3.3 & -12.6 & 393.6 \\ 
H$_3$CSH    (?)&  C$_{1h}$ &  -4.9&   3.1& -11.9&  -4.9 & -13.4 & 472.7 \\ 
HOCl&  C$_{1h}$ &  10.2&  17.8&   6.6&  13.4 &   0.6 & 164.3 \\ 
SO$_2$&  C$_{2v}$ &  22.9&  27.9&  15.4&  20.5 &  -3.8 & 258.5 \\ 
BF$_3$&  D$_{3h}$ &  93.7&  83.5&  79.2&  69.3 &   9.9 & 470.0 \\ 
BCl$_3$&  D$_{3h}$ &  43.7&  59.1&  33.9&  48.3 &  13.1 & 325.0 \\ 
AlF$_3$&  D$_{3h}$ &  43.1&  39.4&  30.1&  26.4 & -17.7 & 425.0 \\ 
AlCl$_3$&  D$_{3h}$ &   3.1&  13.0&  -3.2&   5.5 &   3.3 & 309.0 \\ 
CF$_4$&  Td & 118.0& 105.9& 102.3&  90.1 &  17.2 & 482.0 \\ 
CCl$_4$&  Td &  68.9&  73.5&  58.0&  62.2 &  26.9 & 316.0 \\ 
COS&  C$_{6v}$ &  49.7&  40.7&  41.4&  32.9 &  17.3 & 336.0 \\ 
CS$_2$&  D$_{6h}$ &  39.0&  35.8&  30.9&  28.5 &  23.3 & 280.0 \\ 
COF$_2$&  C$_{2v}$ &  92.2&  79.3&  79.7&  67.3 &  17.4 & 423.0 \\ 
SiF$_4$&  Td &  61.2&  63.7&  44.3&  46.6 &   1.0 & 566.0 \\ 
SiCl$_4$&  Td &  11.3&  12.3&   3.0&   3.2 &  15.9 & 387.0 \\ 
N$_2$O&  C$_{6v}$ &  34.0&  17.4&  28.7&  13.0 &  -4.6 & 270.0 \\ 
C$_2$Cl$_4$&  D$_{2h}$ &  94.2& 101.0&  77.7&  84.7 &  32.8 & 471.0 \\ 
CF$_3$CN&  C$_{3v}$ & 115.6&  98.7&  97.7&  81.4 &  14.0 & 645.0 \\ 
CH$_3$CCH (propyne)&  C$_{3v}$ &  35.0&  27.6&  22.9&  13.5 & -15.0 & 703.0 \\ 
CH$_2$CCH$_2$ (allene)&  D$_{2d}$ &  38.3&  33.4&  24.8&  18.8 & -13.6 & 702.0 \\ 
C$_3$H$_4$ (cyclopropyne)&  C$_{2v}$ &  42.8&  39.1&  30.0&  25.0 &  -2.5 & 679.0 \\ 
CH$_3$CHCH$_2$ (propene)&  C$_{1h}$ &  23.7&  27.6&   9.8&  12.0 & -12.5 & 859.0 \\ 
C$_3$H$_6$&  D$_{3h}$ &  31.9&  36.8&  16.2&  19.8 & -11.4 & 851.0 \\ 
C$_3$H$_8$ (propane)&  C$_{2v}$ &  13.6&  22.0&  -1.9&   4.6 & -16.1 & 1005.0 \\ 
CH$_2$CHCHCH$_2$ (butidene)&  C$_{2v}$ &  48.6&  50.6&  31.5&  31.5 &   1.9 & 1009.0 \\ 
C$_4$H$_6$ (butyne)&  D$_{3d}$ &  50.7&  44.3&  31.5&  24.5 &  -8.8 & 1001.0 \\ 
C$_4$H$_6$ (methylene cylcopropane)&  C$_{2v}$ &  58.3&  58.3&  38.8&  37.1 &  -7.5 & 990.0 \\ 
C$_4$H$_6$ (bicyclobutane)&  C$_{2v}$ &  63.3&  65.3&  43.8&  44.6 &   9.4 & 983.0 \\ 
C$_4$H$_6$ (cyclobutene)&  C$_{2v}$ &  57.9&  61.0&  39.1&  41.2 &   9.2 & 998.0 \\ 
C$_4$H$_8$ (Cyclobutane)&  D$_{4h}$ &  43.9&  52.7&  25.0&  30.9 &   5.4 & 1147.0 \\ 
C$_4$H$_8$ (isobutene)&  C$_{2v}$ &  38.0&  44.4&  18.7&  22.7 & -13.5 & 1156.0 \\ 
C$_4$H$_{10}$ (butane)&  C$_{2h}$ &  26.5&  36.8&   5.8&  14.2 &  -1.7 & 1299.0 \\ 
C$_4$H$_{10}$ (isobutane)&  C$_{3v}$ &  28.4&  37.2&   7.7&  14.7 &  -9.0 & 1301.0 \\ 
C$_5$H$_{8}$ (spiropentane)&  D$_{2d}$ &  79.0&  84.3&  53.2&  57.1 &   8.1 & 1281.0 \\ 
C$_6$H$_{6}$ (benzene)&  D$_{6h}$ & 116.4& 116.3&  90.6&  89.5 &  54.2 & 1362.0 \\ 
CH$_2$F$_{2}$ (difluromethylene)&  C$_{2v}$ &  47.0&  47.0&  36.9&  35.8 & -10.9 & 439.0 \\ 
CHF$_3$ (trifluromethane)&  C$_{3v}$ &  81.8&  75.8&  68.9&  62.3 &  -0.4 & 462.0 \\ 
CH$_2$Cl$_{2}$ (trichloromethane)&  C$_{2v}$ &  26.3&  31.0&  17.6&  21.8 &  -0.0 & 370.0 \\ 
CHCl$_3$ (chloroform)&  C$_{3v}$ &  47.7&  52.4&  37.7&  42.2 &  15.1 & 344.0 \\ 
CH$_3$NH$_2$ (methylamine)&  C$_{1h}$ & -11.9&  -3.9& -19.5& -13.1 & -23.1 & 581.0 \\ 
CH$_3$CN (methyl cynaide)&  C$_{3v}$ &  21.3&  13.7&   9.9&   2.1 & -25.0 & 615.0 \\ 
CH$_3$NO$_{2}$ (nitromethane)&  C$_{1h}$ &  59.0&  55.5&  45.7&  42.1 &   5.5 & 603.0 \\ 
CH$_3$ONO (methyl nitrite)&  C$_{1h}$ &  47.8&  45.3&  36.1&  33.1 &   4.4 & 601.0 \\ 
CH$_3$SiH$_{3}$ (methyl silane)&  C$_{3v}$ & -27.6& -23.1& -33.5& -30.7 & -34.8 & 627.0 \\ 
CHOOH (formic acid)&  C$_{1h}$ &  51.4&  44.9&  40.7&  33.8 &   3.8 & 503.0 \\ 
HCOOCH$_3$ (methyl formate)&  C$_{1h}$ &  61.7&  58.8&  45.4&  41.7 &   1.3 & 788.0 \\ 
CH$_3$CONH$_{2}$ (acetamide)&  C$_{1h}$ &  45.9&  46.9&  29.4&  30.4 &  22.9 & 867.0 \\ 
C$_2$H$_4$NH (aziridine)&  C$_{2v}$ &   5.1&  10.2&  -8.0&  -3.2 &  -9.2 & 719.0 \\ 
CNCN   (cyanogen)&  D$_{6h}$ &  42.3&  18.6&  32.1&   9.7 & -18.2 & 501.0 \\ 
(CH$_3$)$_{2}$NH (dimethylamine)&  C$_{1h}$ &  -3.6&   5.9& -16.7&  -9.0 & -23.4 & 869.0 \\ 
CH$_3$CH$_2$NH$_2$ (trans ethyalmine)&  C$_{1h}$ &  -0.3&   9.2& -13.4&  -5.4 & -20.1 & 877.0 \\ 
CH$_2$CO (ketene)&  C$_{2v}$ &  53.2&  41.6&  41.6&  30.0 &  -1.2 & 533.0 \\ 
C$_2$H$_{4}$O (oxirane)&  C$_{2v}$ &  38.7&  41.5&  25.6&  27.4 &  -4.5 & 651.0 \\ 
CH$_3$CHO   (acetaldehyde)&  C$_{1h}$ &  34.4&  33.9&  21.7&  20.5 &  -8.5 & 677.0 \\ 
HCOCOH (glyoxal)&  C$_{2h}$ &  67.2&  59.1&  53.6&  45.3 &   8.6 & 636.0 \\ 
CH$_3$CH$_2$OH (ethanol)&  C$_{1h}$ &  23.0&  29.3&   9.5&  14.0 & -12.2 & 810.0 \\ 
(CH$_3$)$_{2}$O (dimethylether)&  C$_{2v}$ &  16.7&  23.6&   3.3&   8.2 & -18.5 & 799.0 \\ 
C$_2$H$_4$S (thioxirane)&  C$_{2v}$ &  25.8&  32.2&  13.1&  18.5 &  -0.0 & 624.0 \\ 
(CH$_3$)$_2$SO (dimethyl sulfoxide)&  C$_{1h}$ &  17.2&  30.3&   0.8&  12.0 & -16.0 & 853.0 \\ 
CH$_3$CH$_2$SH (ethanethiol)&  C$_{1h}$ &   8.6&  19.0&  -3.7&   5.3 & -10.3 & 767.0 \\ 
(CH$_3$)$_2$S (dimethyl sulphide)&  C$_{2v}$ &   5.3&  15.1&  -8.2&   0.2 & -11.1 & 766.0 \\ 
CH$_2$CHF (vinyl fluride)&  C$_{1h}$ &  44.0&  45.0&  32.6&  32.5 &  -2.5 & 573.0 \\ 
CH$_3$CH$_2$Cl (ethyl chloride)&  C$_{1h}$ &  19.4&  25.7&   6.8&  12.1 & -12.0 & 691.0 \\ 
CH$_2$CHCl (vinyl chloride)&  C$_{1h}$ &  31.0&  33.7&  20.6&  22.3 &  -1.2 & 542.0 \\ 
CH$_3$CHCN  (acrylonitrile)&  C$_{1h}$ &  44.2&  33.3&  29.7&  18.8 & -12.9 & 761.0 \\ 
(CH$_3$)$_2$CO (acetone)&  C$_{2v}$ &  47.9&  50.3&  29.3&  31.1 &  -6.0 & 978.0 \\ 
CH$_3$COOH (acetic acid)&  C$_{1h}$ &  65.7&  62.5&  48.9&  45.0 &   0.6 & 804.0 \\ 
CH$_3$COF (acetyl fluride)&  C$_{1h}$ &  71.0&  67.8&  55.1&  51.6 &   2.6 & 707.0 \\ 
CH$_3$COCl (acetyl chloride)&  C$_{1h}$ &  57.2&  57.0&  42.8&  42.0 &   3.3 & 669.0 \\ 
CH$_3$CH$_2$CH$_2$Cl (propyl chloride)&  C$_{1h}$ &  32.5&  40.8&  14.6&  21.8 &  -5.4 & 985.0 \\ 
(CH$_3$)$_2$CHOH (isopropanol)&  C$_{1h}$ &  35.0&  43.0&  16.1&  22.6 &  -8.2 & 1108.0 \\ 
CH$_3$CH$_2$OCH$_3$ (methyl ethylether)&  C$_{1h}$ &  32.4&  40.4&  13.3&  19.4 &  -7.6 & 1096.0 \\ 
(CH$_3$)$_3$N (trimethylamine)&  C$_{3v}$ &  15.8&  25.6&  -3.0&   4.6 & -24.4 & 1160.0 \\ 
C$_4$H$_4$O (furan)&  C$_{2v}$ & 103.1& 101.4&  82.3&  79.5 &  39.7 & 992.0 \\ 
C$_4$H$_4$S (thiophene)&  C$_{2v}$ &  87.3&  92.7&  66.0&  71.1 &  51.6 & 960.0 \\ 
C$_4$H$_4$NH (pyrole)&  C$_{2v}$ & -15.1& -12.6& -35.5& -33.9 & -48.0 & 1168.0 \\ 
C$_5$H$_5$N (pyridine)&  C$_{2v}$ & 105.1& 105.4&  80.2&  80.2 &  43.3 & 1234.0 \\ 
HS&  C$_{6v}$ & -12.2&  -9.3& -12.2&  -9.5 &  -9.9 &  87.0 \\ 
CCH  (ethynyl radical)&  C$_{6v}$ &  30.8&  21.8&  24.7&  15.2 &  -1.7 & 262.0 \\ 
CH$_2$CH (vinyl radical)&  C$_{1h}$ &  19.4&  18.7&  11.3&   9.7 &  -8.0 & 443.0 \\ 
CH$_3$CO&  C$_{1h}$ &  43.2&  39.3&  31.4&  27.1 &  -3.5 & 581.0 \\ 
CH$_2$OH (hydroxymethyl)&  C$_{1}$ &  20.2&  21.9&  12.9&  13.3 &  -6.0 & 409.0 \\ 
ClNO&  C$_{6v}$ & -42.6& -44.5& -48.1& -49.6 & -70.4 & 192.0 \\ 
NF$_3$&  C$_{3v}$ &  64.0&  64.3&  57.0&  57.5 &  15.9 & 209.0 \\ 
PF$_3$&  C$_{3v}$ &  41.0&  45.9&  29.6&  34.6 &  29.4 & 359.0 \\ 
O$_3$&  C$_{2v}$ &  50.8&  52.7&  46.7&  48.7 &  34.4 & 148.0 \\ 
F$_2$O&  C$_{2v}$ &  50.6&  55.3&  47.2&  51.3 &  30.7 &  94.0 \\ 
ClF$_3$&  C$_{3v}$ &  -8.2&  -6.9& -14.0& -12.8 & -53.8 & 126.0 \\ 
C$_2$F$_4$&  D$_{2h}$ & 146.2& 138.2& 126.5& 118.6 &  36.5 & 592.0 \\ 
CH$_3$O (methoxy radical)&  C$_{1h}$ &  15.2&  17.2&   7.8&   8.9 &  -8.9 & 400.0 \\ 
CH$_3$CH$_2$O&  C$_{1}$ &  31.0&  35.2&  17.9&  21.3 &  -2.4 & 696.0 \\ 
CH$_3$S (methylsulfide radical)&  C$_{1h}$ &   4.2&   9.4&  -2.6&   1.7 &  -7.0 & 381.0 \\ 
CH$_3$CH$_2$ (ethyl radical)&  C$_{1h}$ &   8.4&  12.9&  -1.3&   1.8 & -16.9 & 601.0 \\ 
(CH$_3$)$_2$CH (isopropanyl radical)&  C$_{1h}$ &  22.6&  28.7&   7.6&  12.2 & -10.9 & 898.0 \\ 
(CH$_3$)$_3$C (isobutanyl radical)&  C$_{3v}$ &  37.3&  45.2&  16.7&  22.8 &  -9.4 & 1195.0 \\ 
NO$_2$ (nitrogen dioxide)&  C$_{2v}$ &  51.5&  41.1&  45.9&  36.2 &  18.0 & 228.0 \\ 
\hline 
 MAE:&  &  33.9&  33.7&  26.4&  25.4 &  14.5 &  - \\
 ME:&   & 25.0&  25.6&  15.5&  15.4 &   -4.8  &- \\
\hline
\hline
\end{longtable}
\endgroup

\LTchunksize=60
\begingroup
\squeezetable
\begin{longtable}{ldcccc}
\caption{The comparison of bond distances of  selected molecules 
 from the G2-148 set, calculated within different computational 
models- 
 M1:  SR-$\alpha_{EA}$/6311G**/RIJ. 
 M2:  SR-$\alpha_{EA}$/DGDZVP2/A2, 
 M3:  SR-$\alpha_{HF}$/6311G**/RIJ, 
 M4:  SR-$\alpha_{HF}$/DGDZVP2/A2, 
All distances are in \AA.  The mean absolute error (MAE) and mean error  
are given in last two rows. The experimental bond lengths 
are from Ref. \onlinecite{TVS98}}
\label{tab:BL}
\\
\hline\hline
 Molecule         &    ~M1~ & ~M2~ & ~M3~ & ~M4~ & ~{\em Exact} \\
\hline 
H$_2$& 0.746& 0.742& 0.747& 0.743 & 0.741\\ 
LiH& 1.562& 1.628& 1.573& 1.631 & 1.595\\ 
BeH& 1.309& 1.412& 1.320& 1.425 & 1.343\\ 
CH& 1.113& 1.117& 1.117& 1.125 & 1.120\\ 
NH& 1.033& 1.037& 1.035& 1.042 & 1.036\\ 
OH& 0.960& 0.967& 0.961& 0.970 & 0.971\\ 
HF& 0.905& 0.912& 0.903& 0.914 & 0.917\\ 
Li$_2$& 2.666& 2.722& 2.690& 2.750 & 2.673\\ 
LiF& 1.513& 1.596& 1.520& 1.610 & 1.564\\ 
CN& 1.142& 1.159& 1.150& 1.167 & 1.172\\ 
CO& 1.105& 1.127& 1.112& 1.134 & 1.128\\ 
N$_2$& 1.077& 1.097& 1.084& 1.103 & 1.098\\ 
NO& 1.125& 1.148& 1.133& 1.156 & 1.151\\ 
O$_2$& 1.175& 1.197& 1.186& 1.207 & 1.208\\ 
F$_2$& 1.356& 1.362& 1.371& 1.376 & 1.412\\ 
HCl& 1.269& 1.266& 1.269& 1.269 & 1.275\\ 
ClF& 1.628& 1.611& 1.643& 1.626 & 1.628\\ 
ClO& 1.559& 1.550& 1.575& 1.565 & 1.570\\ 
ClO& 1.559& 1.550& 1.575& 1.565 & 1.570\\ 
Na$_2$& 2.990& 3.000& 3.015& 3.024 & 3.079\\ 
NaCl& 2.366& 2.324& 2.383& 2.338 & 2.361\\ 
NH& 1.033& 1.037& 1.035& 1.042 & 1.036\\ 
P$_2$& 1.876& 1.886& 1.886& 1.896 & 1.893\\ 
S$_2$& 1.893& 1.897& 1.905& 1.909 & 1.889\\ 
CS& 1.511& 1.518& 1.523& 1.529 & 1.535\\ 
SiO& 1.501& 1.507& 1.510& 1.516 & 1.510\\ 
SO& 1.487& 1.486& 1.498& 1.496 & 1.481\\ 
\hline
 MAE:& 0.020& 0.018& 0.016& 0.017 & - \\ 
 MEAN:& -0.018& -0.004& -0.009& 0.006 & - \\ 
\hline
\vspace{5in}
\end{longtable}
\endgroup

\LTchunksize=75
\begingroup
\squeezetable
\begin{longtable}{ldcccc}
\caption{The deviation in the first vertical ionization energies with respect to 
experimental ionization potentials  for the subsets G2 set 
of molecules (49 in total) within different computational models- 
 M1:  SR-$\alpha_{EA}$/6311G**/RIJ. 
 M2:  SR-$\alpha_{EA}$/DGDZVP2/A2, 
 M3:  SR-$\alpha_{HF}$/6311G**/RIJ, 
 M4:  SR-$\alpha_{HF}$/DGDZVP2/A2, 
All energies are in eV and are calculated at the optimized geometries 
of molecules in the respective model. The last column contains exact ionization potentials; 
these  are obtained from tabulation in Ref. \onlinecite{Curtiss_IP} }
\label{tab:IP}
\\
\hline\hline
 Molecule         &   ~ M1~ & ~M2~ & ~M3~ & ~M4~ & ~{\em Exact}~ \\
\hline 
\hline 
 B&   0.6&   0.6&   0.3&   0.4 &   8.3\\ 
Be&   0.1&   0.1&  -0.1&  -0.1 &   9.3\\ 
 C&   0.8&   1.0&   0.5&   0.6 &  11.3\\ 
 N&   1.2&   1.2&   0.8&   0.8 &  14.5\\ 
 O&  -0.5&  -0.4&  -0.8&  -0.7 &  13.6\\ 
 f&   0.2&   0.3&  -0.2&  -0.1 &  17.4\\ 
Na&   0.1&   0.1&   0.0&   0.0 &   5.1\\ 
Mg&  -0.0&   0.0&  -0.1&  -0.1 &   7.7\\ 
Al&  -0.2&  -0.2&  -0.4&  -0.3 &   6.0\\ 
Si&  -0.1&  -0.1&  -0.3&  -0.2 &   8.2\\ 
 P&   0.0&   0.1&  -0.2&  -0.1 &  10.5\\ 
 S&  -0.9&  -0.8&  -1.1&  -1.0 &  10.4\\ 
Cl&  -0.6&  -0.4&  -0.9&  -0.7 &  13.0\\ 
CH$_4$&   1.5&   1.5&   1.4&   1.4 &  12.6\\ 
NH$_3$&   0.4&   0.5&   0.2&   0.3 &  10.2\\ 
OH&  -0.4&  -0.2&  -0.6&  -0.5 &  13.0\\ 
H$_2$O&  -0.0&   0.2&  -0.2&  -0.1 &  12.6\\ 
HF&   0.4&   0.6&   0.1&   0.3 &  16.0\\ 
SiH$_4$&   1.0&   1.0&   1.0&   0.9 &  11.0\\ 
PH$_2$&  -0.0&   0.0&  -0.2&  -0.1 &   9.8\\ 
PH$_3$&   0.4&   0.4&   0.3&   0.3 &   9.9\\ 
HS&  -0.7&  -0.6&  -0.9&  -0.8 &  10.4\\ 
HCl&  -0.4&  -0.3&  -0.6&  -0.5 &  12.8\\ 
CO&  -0.5&  -0.3&  -0.7&  -0.6 &  14.0\\ 
O$_2$&   1.0&   0.9&   0.5&   0.5 &  12.1\\ 
P$_2$&  -0.3&  -0.2&  -0.5&  -0.4 &  10.5\\ 
S$_2$&  -0.2&  -0.0&  -0.4&  -0.3 &   9.4\\ 
Cl$_2$&  -0.9&  -0.6&  -1.1&  -0.9 &  11.5\\ 
ClF&  -0.5&  -0.3&  -0.8&  -0.6 &  12.7\\ 
CS&  -0.7&  -0.6&  -0.9&  -0.8 &  11.3\\ 
BF$_3$&  -0.4&  -0.2&  -0.9&  -0.7 &  15.6\\ 
BCl$_3$&  -1.0&  -0.9&  -1.2&  -1.2 &  11.6\\ 
CO$_2$&   0.3&   0.4&  -0.1&   0.0 &  13.8\\ 
CS$_2$&  -0.1&  -0.0&  -0.3&  -0.2 &  10.1\\ 
CH$_3$&   0.4&   0.4&   0.3&   0.3 &   9.8\\ 
CN&   0.9&   1.0&   0.6&   0.7 &  13.6\\ 
CH$_3$O (methoxy radical)&   0.2&   0.3&   0.0&   0.1 &  10.7\\ 
H$_3$CO$_{}$H &  -0.2&  -0.2&  -0.4&  -0.3 &  10.8\\ 
CH$_2$OH (hydroxymethyl)&   0.4&   0.5&   0.2&   0.2 &   7.5\\ 
CH$_2$($^1A_1$)&   0.7&   0.8&   0.6&   0.6 &   9.4\\ 
CH$_3$Cl &  -0.4&  -0.3&  -0.5&  -0.5 &  11.2\\ 
CNCN   (cyanogen)&   0.4&   0.4&   0.0&   0.1 &  13.4\\ 
C$_4$H$_4$O (furan)&   0.6&   0.6&   0.4&   0.4 &   8.8\\ 
NH&  -0.8&  -0.8&  -1.0&  -0.9 &  13.5\\ 
NH$_2$&   0.3&   0.3&   0.1&   0.2 &  11.1\\ 
SiH$_3$&  -0.2&  -0.2&  -0.3&  -0.3 &   8.1\\ 
C$_6$H$_{6}$ (benzene)&   0.4&   0.4&   0.3&   0.2 &   9.2\\ 
Si$_2$H$_6$&   0.5&   0.4&   0.4&   0.3 &   9.7\\ 
PH$_2$&  -0.0&   0.0&  -0.2&  -0.1 &   9.8\\ 
\hline
 MAE:&   0.5&   0.5&   0.5&   0.5& - \\
 Mean:&   0.0&   0.1&  -0.2&  -0.1& - \\
\hline
\end{longtable}
\endgroup

\end{document}